\documentclass[preprint]{aa}

\usepackage{txfonts}
\usepackage{natbib}
\usepackage{units}
\usepackage{graphicx}
\usepackage{epsfig}
\usepackage{url}
\bibpunct{(}{)}{;}{a}{}{,}

\newcommand{\bea}{\begin{eqnarray*}}
\newcommand{\eea}{\end{eqnarray*}}
\newcommand{\bean}{\begin{eqnarray}}
\newcommand{\eean}{\end{eqnarray}}

\newcommand{\kms}{\ensuremath{\mathrm{km}\,\mathrm{s}^{-1}}}
\def\specchar#1{\uppercase{#1}}    

\newcommand\mgiihk{\mbox{Mg\,\specchar{ii}\,h\,\&\,k}} 
\newcommand\eg{e.g.,}              
\newcommand{\bifrost}{{\textsl{Bifrost}}}
\newcommand{\iris}{{IRIS}}
\newcommand{\edt}[1]{{ #1}}

\DeclareRobustCommand\sfrac[1]{\@ifnextchar/{\@sfrac{#1}}%
                                            {\@sfrac{#1}/}}
\def\@sfrac#1/#2{\leavevmode\kern.1em\raise.5ex
         \hbox{$\m@th\mbox{\fontsize\sf@size\z@
                           \selectfont#1}$}\kern-.1em
         /\kern-.15em\lower.25ex
          \hbox{$\m@th\mbox{\fontsize\sf@size\z@
                            \selectfont#2}$}}

\title{A publicly available simulation of an enhanced network region of the Sun}
\author{Mats Carlsson\inst{1}
\and Viggo H. Hansteen\inst{1}
\and Boris  V. Gudiksen\inst{1}
\and Jorrit Leenaarts\inst{2}
\and Bart De Pontieu\inst{3,1}
}

\institute{
Institute of Theoretical Astrophysics, University of Oslo, P.O. Box
1029 Blindern, N-0315 Oslo, Norway
\and
Institute for Solar Physics, Department of Astronomy,
  Stockholm University,
AlbaNova University Centre, SE-106 91 Stockholm Sweden
\and
Lockheed Martin Solar \& Astrophysics Lab,
         Org.\ A021S, Bldg.\ 252, 3251 Hanover Street
         Palo Alto, CA~94304 USA}

\date{\today / -  }

\authorrunning{Carlsson et al.}
\titlerunning{Simulations of enhanced network}
\abstract
{The solar chromosphere is
the interface between the solar surface and the solar corona. 
Modelling of this region is difficult because it represents the transition from optically thick to thin radiation 
escape, from gas-pressure domination to magnetic-pressure domination, 
from a neutral to an ionised state, from MHD to plasma physics, and from near-equilibrium (LTE) to non-equilibrium conditions.} 
{Our aim is to provide the community with realistic simulations of the magnetic solar outer atmosphere. 
This will enable detailed
comparison of existing and upcoming observations with synthetic observables from the simulations, 
thereby elucidating the complex interactions of magnetic fields and plasma that are crucial for our
understanding of the dynamic outer atmosphere.} 
{We used the radiation magnetohydrodynamics code \bifrost\ to perform simulations of a computational volume
with a magnetic field topology similar to an enhanced network area on the Sun.} 
{The full simulation cubes are made available online. 
The general properties of the simulation are discussed, and limitations are discussed.} 
{} 
\begin{document}

\maketitle
 
\section{Introduction}\label{sec:introduction}
The structure and dynamics of the outer solar atmosphere are set by
magnetism. In the convection zone, the gas pressure exceeds the 
magnetic pressure in all but the strongest magnetic flux
concentrations and the field is moved around by the plasma. These
motions drive flows of energy and mass through the chromosphere into
the corona. Most of the energy that is transported to the outer solar
atmosphere through work done on the magnetic fields is 
radiated away in the chromosphere. 
It is also in the chromosphere that the dynamics change from
gas-pressure-dominated behaviour to magnetic
force dominance. The layer where the sound speed is equal to the Alfv\'en speed is located in the chromosphere, and conversion between different wave modes may occur.
The ionization state goes
from almost neutral to full ionization in the corona.  The radiation
goes from optically thick to optically thin, from local thermodynamic
equilibrium (LTE) to non-equilibrium conditions. All these
transitions make chromospheric physics very complicated, 
and the chromosphere may be the least understood
region of 
the Sun \citep{Judge+Peter1998}.

An early class of models of the solar chromosphere were
1D, semi-empirical models. Only in 1D was it possible to
solve the non-LTE radiative transfer equations needed to produce
synthetic observables that could be compared with observations. Since
the energy transportation and dissipation mechanisms responsible for
heating the chromosphere were unknown, the energy equation was
replaced by treating the temperature as a function of height as a free
parameter. Early reference models of this kind were the
Bilderberg continuum atmosphere \citep[BCA,][]{1968SoPh....3....5G} and the
Harvard-Smithsonian reference atmosphere \citep[HSRA,][]{1971SoPh...18..347G}.
With increased amounts of observables through continuum observations
in the UV from Skylab, a series of models for
six different components of the quiet solar chromosphere were constructed in a
seminal series of papers \citep{1973ApJ...184..605V,
 1976ApJS...30....1V, 1981ApJS...45..635V}, and the model corresponding
most closely to the average quiet solar chromosphere, often denoted
VAL3C, is the most cited solar chromospheric model. 
Later models have
improved the fit in the temperature-minimum region
\citep{1985cdm..proc...67A, 1986ApJ...306..284M}
and removed
the need for a temperature plateau to reproduce the hydrogen
Lyman-$\alpha$ line
\citep{1990ApJ...355..700F, 1991ApJ...377..712F, 1993ApJ...406..319F}.
See \citet{2002JAD.....8....8R} for an overview of 1D
solar model atmospheres.

These models have been (and still are) very useful in providing model
atmospheres with reasonable chromospheric conditions, and they can be used
as numerical laboratories for exploring chromospheric
line formation. It is important to keep in mind, though, that many
different atmospheric models are consistent with a certain set of
observables; \citet{1995ApJ...440L..29C} showed that a dynamic
atmosphere with strong shocks gave the same temporal average UV
continuum intensities as a VAL type model even though the average
temperature structure was close to the radiative equilibrium solution.

Instead of using a trial-and-error way of adjusting the temperature
structure, it is in principle possible to formulate an inversion strategy whereby
the ``best'' model is arrived at through a formal definition of a
``norm'' and an automatic algorithm to minimise this norm. This can
even be done in 3D including effects of an observational
point spread function and the effects of 3D scattering
\citep[e.g.,][]{2012A&A...548A...5V,2015A&A...577A...7S,2015A&A...577A.140A}.
However, a fully
unconstrained approach is ill-conditioned because there are more free
parameters than observables. It is therefore crucial to develop proper
strategies to arrive at physically motivated constraints in inversions.

The purpose of this paper is to describe a ``realistic'' numerical simulation of the
solar outer atmosphere, extending from the upper convection zone to the corona, that is not determined by any fitting procedure to
observations. By ``realistic'' we mean that we have gone to great
lengths in trying to include the relevant physical processes in the
numerical code and minimise the number of free
parameters. Observations have thus NOT gone into constraining the
model and a comparison of synthetic observables with observations will
give information on what physics is missing in the numerical
simulation. We also believe that the simulation sequence is very
useful as a numerical laboratory to determine how observables depend
on the atmospheric parameters. This will be true even if the models
fail to reproduce certain detailed observations (which we already know
is the case) as long as the model includes much of the important
physics. For this reason it is important to have an understanding of
what physics went into the model, what the approximations are,
the general properties of the model and how they can be/cannot be
used. This paper aims at providing such a detailed description.

A number of papers have used the simulation sequence described here.
The formation of the H-$\alpha$ line was studied in
\citet{2012ApJ...749..136L}, the Hanl\'e effect of Ly-$\alpha$ was
studied by \citet{2012ApJ...758L..43S,2015ApJ...803...65S}
and the signatures of heating
of the magnetic chromosphere were treated by
\citet{2013ApJ...764L..11D}. 
\citet{2015A&A...575A..15L}
studied the diagnosing of the chromospheric thermal structure
using millimeter radiation and 
\citet{2015ApJ...802..136L} studied the nature of H-$\alpha$ fibrils in the solar chromosphere.

With the advent of chromospheric observations with high spatial and temporal resolution from
the  NASA
Small Explorer satellite Interface Region Imaging Spectrograph 
\citep[\iris,][]{2014SoPh..289.2733D}, we feel that it is crucial to use detailed 
numerical simulations, like the one presented here, in order to improve
our understanding of what the observations tell us. This was also the
initial driver for making this simulation sequence publicly available.
A series of papers have been devoted to the
formation of lines that are observable with \iris:
\citet{2013ApJ...772...89L, 2013ApJ...772...90L,2013ApJ...778..143P}
treated the formation of the \ion{Mg}{ii} h \& k lines,
\citet{2015ApJ...806...14P} treated the formation of the \ion{Mg}{ii} UV subordinate lines,
\citet{bhavna1,bhavna2} treated the formation of the \ion{C}{ii} multiplet near 133.5\,nm
and 
\citet{lin1} showed that the \ion{O}{i} line at 135.56\,nm is an excellent diagnostic
of non-thermal velocities in the solar chromosphere. All these papers used
snapshots from the
current simulation as a laboratory for exploring line formation
characteristics and relations between atmospheric conditions and
observables. 

The structure of this paper is as follows: In Section~\ref{sec:bifrost} we give a short description 
of the \bifrost\ code, in Section~\ref{sec:simulation} we describe the general properties of 
the simulation sequence and in Section~\ref{sec:data} we describe the data format and
how to access the simulation data and we end with Discussion and Conclusions in Section~\ref{sec:discussion}.

\section{Bifrost}\label{sec:bifrost}
The simulation described here has been performed with the
3D Radiation Magneto-Hydrodynamic (RMHD) code
\bifrost.  \bifrost\ is a general, flexible and massively parallel code
described in detail in 
\citet{2011A&A...531A.154G}. 
In short, \bifrost\ solves the MHD equations on a staggered grid
using a 5th/6th order compact finite difference scheme. The effects of
radiation in the energy balance are \edt{taken into account} by solving
the radiative transfer equations along rays through the computational
domain using a short-characteristic method \edt{and} multi-group opacities
\citep{1982A&A...107....1N} 
\edt{with four opacity groups}
modified to take into account scattering 
\citep{2000ApJ...536..465S}. 
See \citet{2010A&A...517A..49H} 
for a detailed description of the
treatment of the radiative transfer. Chromospheric radiative losses
are calculated in non-LTE using simplified recipes 
\citep{2012A&A...539A..39C} 
based on detailed 1D full non-LTE radiative transfer
simulations using the RADYN code
\citep{1992ApJ...397L..59C,1995ApJ...440L..29C,1997ApJ...481..500C,2002ApJ...572..626C}.
Optically thin radiative losses are taken into account
using tables calculated from atomic data in CHIANTI, version 5
\citep{Chianti1,Chianti5}.  Thermal conduction
becomes important at high temperatures and is included using operator
splitting with an implicit formulation based on a multi-grid method.
\edt{\bifrost\ is an explicit code with diffusive terms in the equations
in order to ensure stability. The diffusive operator employed is split in
a small global diffusive term and a location specific hyper diffusion term,
see \citet{2011A&A...531A.154G} for details. In this simulation we do not
include any terms taking care of ambipolar diffusion or Hall currents.}
\bifrost\ is a very general modeling code and a variety of
modules are available for boundary conditions and the
equation of state. For the simulation described in this paper we have
included non-equilibrium ionization of hydrogen following the
description by 
\citet{2007A&A...473..625L} 
based on the approximations by 
\citet{Sollum1999}.
The background opacities are given by the old Uppsala background
opacity package 
\citep{Gustafsson1973} 
\edt{and abundances are from \citet{1975A&A....42..407G}}.


\section{Simulation}\label{sec:simulation}
We have set up our simulation with the aim of studying processes in
the solar chromosphere with a magnetic field configuration that we
characterize as ``enhanced network''. 
The computational box is 24 by 24 Mm$^2$ horizontally with periodic
boundary conditions and extends 2.4 Mm below the visible surface
(defined as the average height where optical depth at 500\,nm is unity;
\edt{this is also the zero point of our height scale.}) and
14.4\,Mm above encompassing the upper part of the convection zone, the
photosphere, chromosphere, transition region and corona. The
computational box is 504x504x496 grid points giving 48 km resolution
horizontally and a variable grid separation in the vertical direction
varying from 19 km in the photosphere and chromosphere up to 5 Mm
height and then increasing to 100 km at the top boundary. Both the top
and bottom boundaries are
 transparent. 
\edt{The top boundary is implemented using the characteristic equations
\citep{2011A&A...531A.154G}. At the bottom boundary, the magnetic field is
passively advected with no extra field fed into the computational domain.}

The simulation was initialised from a hydrodynamic simulation of size
6x6x3\,Mm$^3$ that had reached a relaxed state. This simulation reached
2.4\,Mm below the visible surface but only 0.5\,Mm above. The simulation
was expanded horizontally (since it is periodic horizontally this just
entails replicating the numerical domain to the larger size), first to
12x12\,Mm$^2$ and then to 24x24\,Mm$^2$. At each step 
\edt{small random perturbations were introduced and}
the simulation was
run long enough that the horizontal periodicity from the startup
vanished. The photospheric simulation was run for ten hours of solar time
at a size of 24x24x3\,Mm$^3$. This relaxed hydrodynamic state was then
expanded to 24x24x17\,Mm$^3$ by adding a chromosphere and corona in 
hydrostatic equilibrium and a temperature structure taken from a previous
simulation. This temperature structure was just an initial condition and the
temperature was not fixed in the following evolution.
The time of the addition of the chromosphere and corona is taken as $t=0$ in the simulation. 
This state was allowed to relax for 1750\,s
of solar time to get rid of inconsistencies in the lower chromosphere. Because
there is still no magnetic field, the upper part slowly cools and at $t=1750$\,s the
temperature at the upper boundary is 250~kK. At that point in time a 
large scale magnetic field was added. 

\begin{figure}[htbp]
\includegraphics[width=\columnwidth]{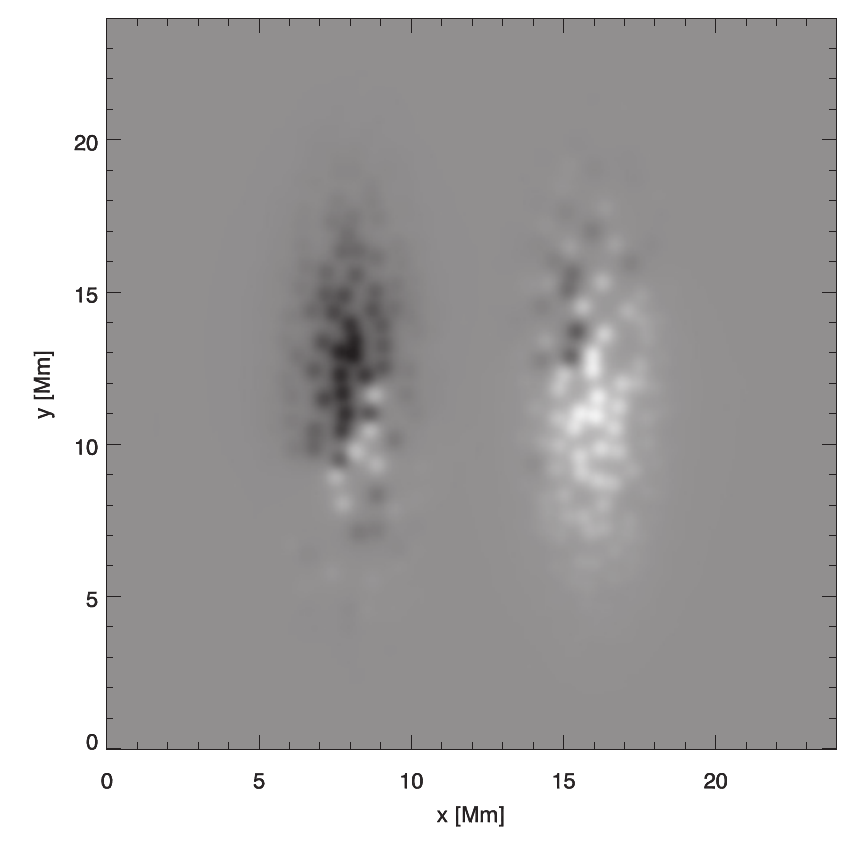}
\caption{Initial vertical magnetic field at the bottom of the computational domain. The maximum
magnetic field strength is 0.8\,kG (both in the bottom plane and the colour bar range)
and the average signed field strength is close to zero (0.025\,G).
\label{fig:bottomBz}}
\end{figure}

The magnetic field was added by specifying the vertical field at the
bottom of the computational domain with a potential field
extrapolation into the rest of the domain. The field at the bottom
boundary was specified to have two patches of opposite dominant
polarity separated by 8\,Mm, with an overall balanced flux, see Figure~\ref{fig:bottomBz}.  
The magnetic field is very quickly
swept to the downdrafts of the convective pattern and the potential
character is also quickly lost in the upper part of the simulation. The moving
around of the magnetic field by convection gives a Poynting flux into
the upper part of the simulation. This magnetic energy is dissipated
and creates a chromosphere and corona.  
The simulation was \edt{run 
assuming instantaneous hydrogen ionization
equilibrium until the non-equilibrium hydrogen ionization was
switched on at $t=3020$\,s.} The simulation was stopped at $t=5440$\,s.
The first published snapshot is snapshot 385 at $t=3850$\,s. 
\edt{The various steps are given in Table~\ref{table:times}}

\begin{table}[htbp]
\caption{Simulation timeline}
\label{table:times}
\begin{tabular}{r l}
\hline\hline
time & event \\
(s) & \\
\hline
-36000 & 24x24x3 Mm$^3$ simulation extending to z=0.5 Mm\\
0 & region added extending to z=14.4\,Mm\\
1750 & magnetic field added \\
3020 & non-equilibrium hydrogen ionization switched on \\
3850 & first published snapshot \\
5440 & last published snapshot \\
\hline
\end{tabular}
\end{table}

\subsection{Magnetic field}\label{subsec:magfield}
The computational box is too small to allow for the build-up of a magnetic field from global
dynamo action. The field inserted at the bottom boundary, as described above, is the
main free parameter of the simulation. It is therefore important to characterize 
the magnetic field that results from the continuous processing of the initial magnetic field by
the convection.

\begin{figure}[htbp]
\includegraphics[width=\columnwidth]{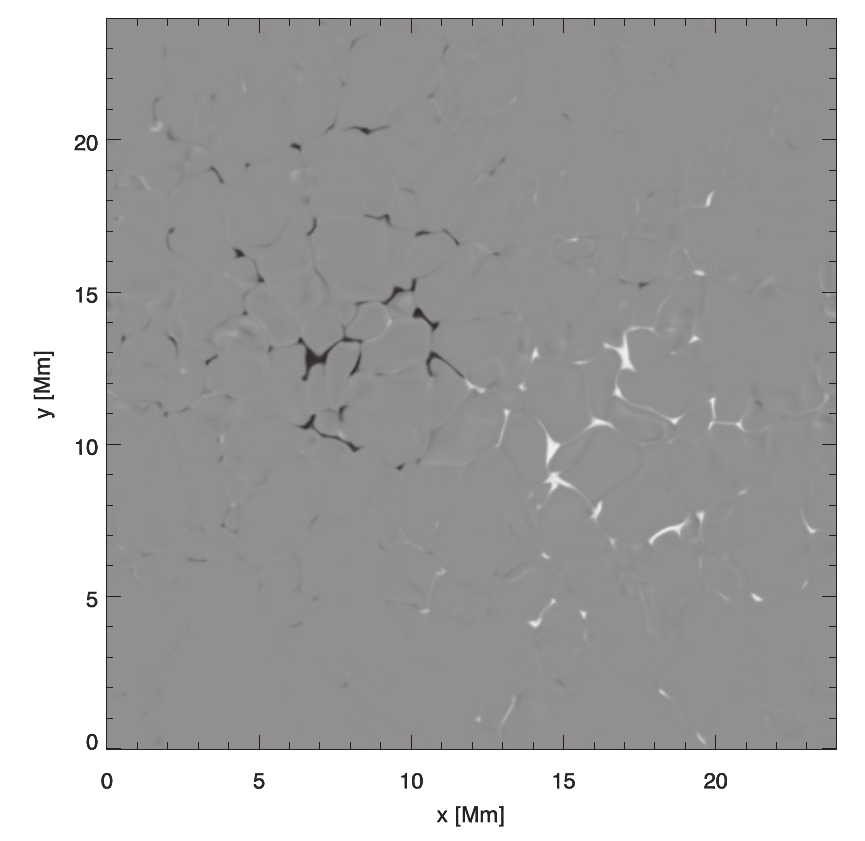}
\caption{Vertical magnetic field strength at $z=0$ and $t=3850$\,s. The 
field has been swept to the intergranular lanes. The maximum field-strength 
is 1.9\,kG. The colour bar range is [-2\,kG, 2\,kG].
\label{fig:Bz0}}
\end{figure}

The average unsigned magnetic field strength in the
photosphere is 48\,G (5\,mT). The vertical magnetic field at $z=0$ at
$t=3850$\,s (the snapshot used for most publications in the list in
Section~\ref{sec:introduction})
is shown in Figure~\ref{fig:Bz0}. The field has been swept to the intergranular lanes but
the initial two dominant polarity patches separated by 8\,Mm are still seen. 

\begin{figure}[htbp]
\includegraphics[width=\columnwidth]{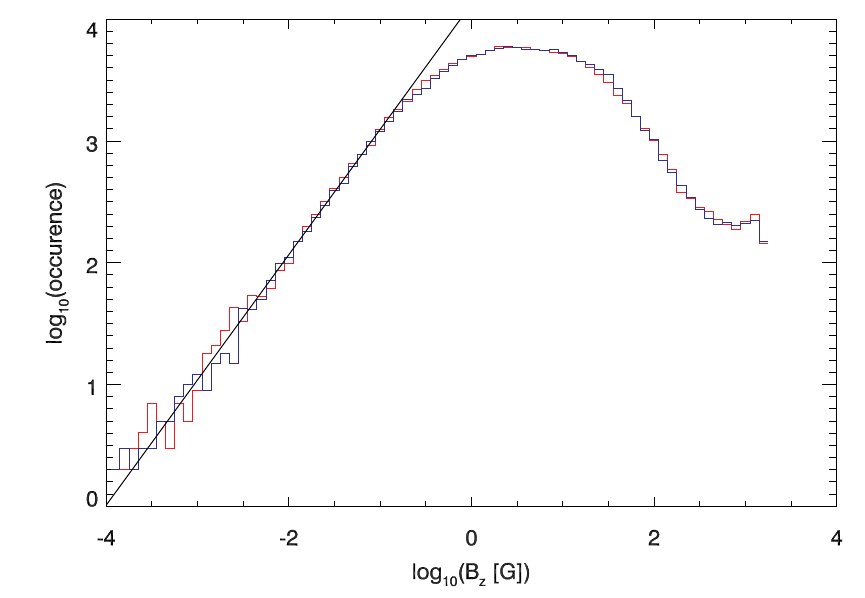}
\caption{Histogram of the vertical magnetic field strength, $B_z$ at $z=0$ and $t=3850$\,s for positive $B_z$ (red)
and negative $B_z$ (blue). The straight black line shows a fit to the field with a strength below 0.1\,G. The slope is 1.03 in the log-log plot.
\label{fig:Bzhist}}
\end{figure}

We can further characterize the magnetic field by the distribution of field-strengths. Figure~\ref{fig:Bzhist} shows a histogram of the
vertical magnetic field strength at $z=0$ at $t=3850$\,s. There is no difference between the distributions of positive and negative $B_z$.
The weaker field follows a power law distribution with a slope of one in the log-log diagram. The magnetic field distribution does not change
significantly during the simulation timespan of 2420\,s.

\begin{figure}[htbp]
\includegraphics[width=\columnwidth]{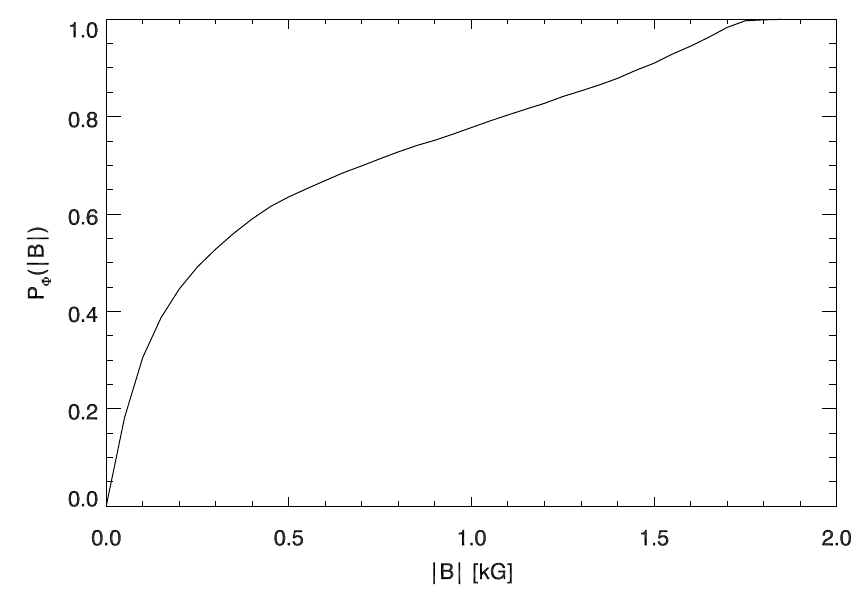}
\caption{\edt{Flux-based probability distribution at $z=0$ and $t=3850$\,s showing the fraction of the total absolute flux that has a flux density of less than $|$B$|$. }\label{fig:pphi}}
\end{figure}

\edt{The flux-based probability distribution \citep{2003A&A...406.1083S} is shown in Figure~\ref{fig:pphi}. This probability distribution shows the 
fraction of the total absolute flux that has a flux density of less than a given value. From the figure we can see that 78\% of the absolute flux at
$z=0$ is in areas with a flux density less than 1\,kG.}

\begin{figure}[htbp]
\includegraphics[width=\columnwidth]{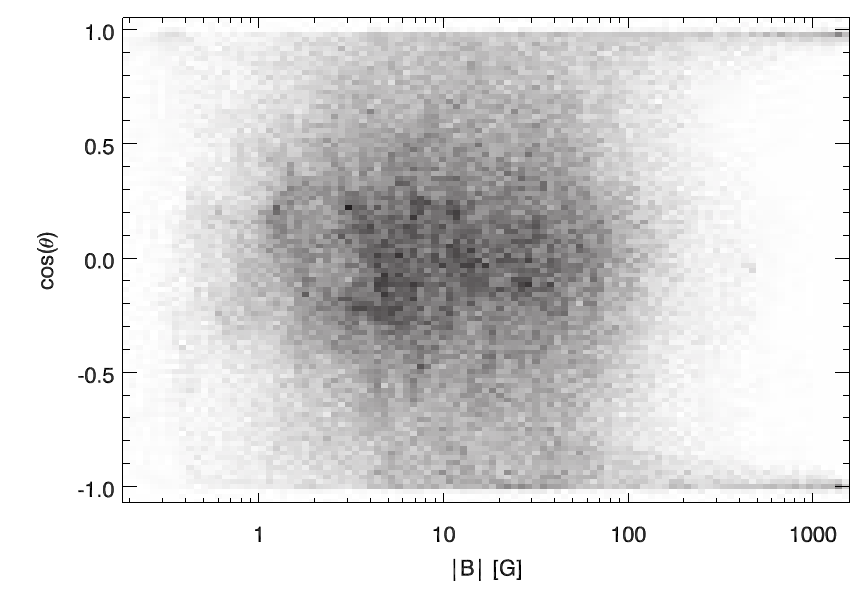}
\caption{Joint probability distribution function (JPDF) of the cosine of the magnetic field 
angle to the vertical and the magnetic field strength at $z=0$ and $t=3850$\,s. 
\label{fig:Bangle}}
\end{figure}

The distribution of field angles at $z=0$ is shown in Figure~\ref{fig:Bangle}. Most of the field is
pretty horizontal $|\cos(\theta)|<0.3$ but the strongest field ($|B|>300$\,G) is vertical.

\begin{figure}[htbp]
\includegraphics[width=\columnwidth]{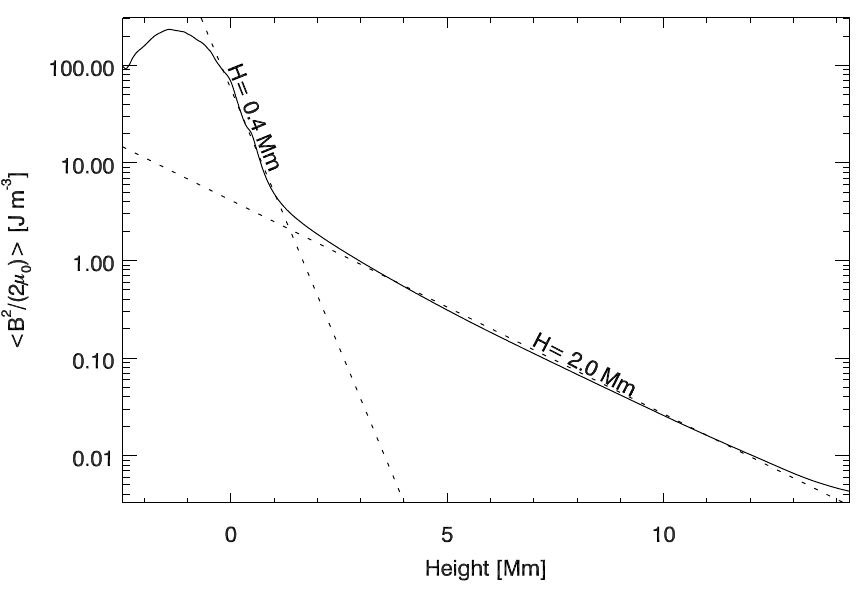}
\caption{Horizontally averaged magnetic energy density ($B^2/(2\mu_0)$) as function of height for the snapshot at $t=3850$\,s.
The scaleheight in the lower chromosphere ($z=0.2-1.2$\,Mm) is 0.4\,Mm and in the upper chromosphere and corona it is roughly 2.0\,Mm
(dotted lines).
\label{fig:b2}}
\end{figure}

The Joule heating in the simulation at heights from the upper photosphere to the corona scales roughly with the magnetic energy
density, $B^2/(2\mu_0)$ 
\citep{2010ApJ...718.1070H,2005ApJ...618.1020G}. 
Figure~\ref{fig:b2} shows the horizontally averaged magnetic energy density as a function of height in the snapshot at $t=3850$\,s
(it is very similar in other snapshots). The scaleheight of the magnetic energy density is about 0.4\,Mm in the lower chromosphere 
($z=0.2-1.2$\,Mm) and increases to about 2\,Mm scaleheight in the upper chromosphere and corona.

\subsection{Photosphere}\label{subsec:photosphere}
\begin{figure}[htbp]
\includegraphics[width=\columnwidth]{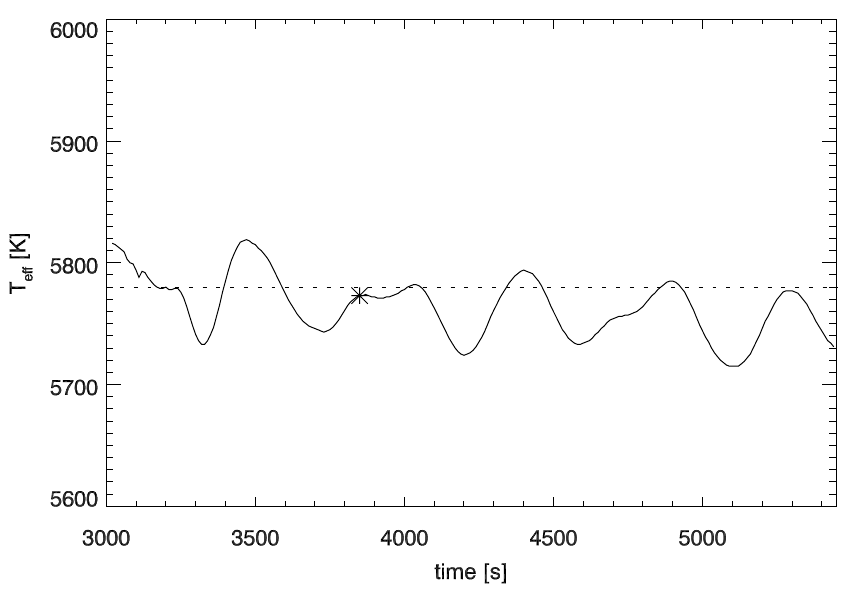}
\caption{Effective temperature as a function of time. The solar effective temperature of 5780\,K is marked with a dotted line and
the much used snapshot at $t=3850$\,s (also the first published snapshot) with a star.
\label{fig:teff}}
\end{figure}

The focus of the current simulation is the chromosphere and corona; for photospheric studies there are other simulations
available with better numerical resolution, better description of the photospheric radiative transfer (\eg\ more opacity bins) and 
more modern continuum opacity data 
\citep[\eg\ with the codes CO$^5$BOLD, MURaM and Stagger, see][for a comparison of the codes]{2012A&A...539A.121B}. 
Our choice of background opacities from the old Uppsala package
\citep{Gustafsson1973} 
was motivated by the availability of a well relaxed hydrodynamical model but is not the ideal choice if the aim is a detailed
comparison of photospheric observables. It is also important to keep in mind that the effective temperature of the simulation 
is not set directly but only indirectly from specifying the incoming entropy at the lower boundary. The effective temperature thus
varies in time with possible drifts with rather long timescales (set by the typical timescales at the bottom boundary).
Figure~\ref{fig:teff} shows the temporal variation of the effective temperature of the simulation. Oscillations with periods of 
350--500\,s are seen 
(for a closer analysis, see Section~\ref{subsec:oscillations}) as well as a downward secular trend. 
The often used snapshot at $t=3850$\,s has an effective temperature
of 5773\,K, close to the solar value.

\subsection{Oscillations}\label{subsec:oscillations}

\begin{figure}[htbp]
\includegraphics[width=\columnwidth]{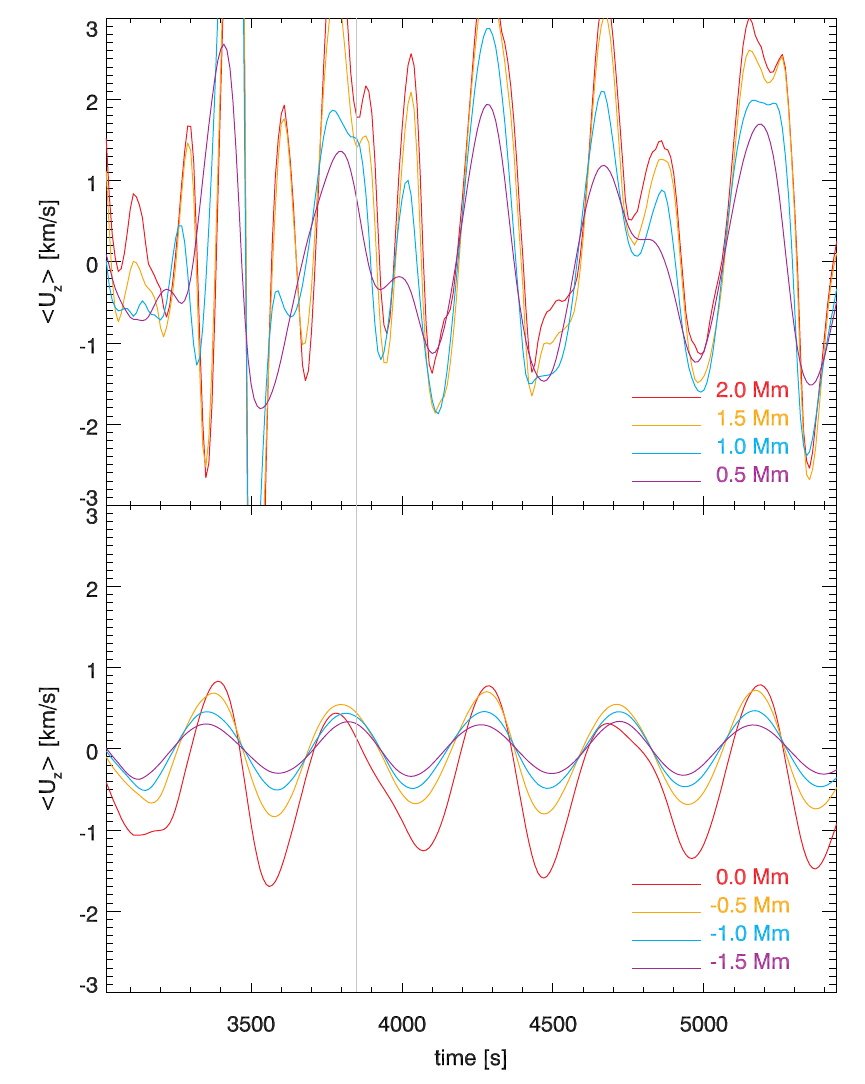}
\caption{Horizontally averaged vertical velocity (positive is downflow) as function of time for eight heights 
(four in the upper panel and four in the lower panel, heights as given in the legend). The start of the published 
sequence of snapshots is indicated at $t=3850$\,s as a grey line.
\label{fig:osc}}
\end{figure}

As is obvious in Figure~\ref{fig:teff}, there are oscillations in the simulation box. 
The lower boundary is a pressure node 
reflecting acoustic waves to mimic the refraction of acoustic waves in the solar deeper atmosphere. 
The excitation of p-modes is similar to the real Sun but the energy is spread over a very limited set of 
modes giving them much larger amplitude (especially the global mode) compared with the Sun
\citep{2001ApJ...546..585S}.
The horizontally averaged
vertical velocity at eight heights, ranging from $z=-1.5$\,Mm to $z=2.0$\,Mm, is given in Figure~\ref{fig:osc}.
The average velocity in the photosphere (lower panel) is dominated by global oscillations that are in phase with 
a period of 450\,s. At $z=0$, the average velocity is negative (upward) because the lower density in the granular upflows
than in the intergranular downflows give a negative average velocity for a zero average massflux.
At $z=0$ the amplitude of the oscillations is about 1\,\kms. In the chromosphere (upper panel) we have a mixture of 
the global oscillations and propagating waves. 

\begin{figure}[htbp]
\includegraphics[width=\columnwidth]{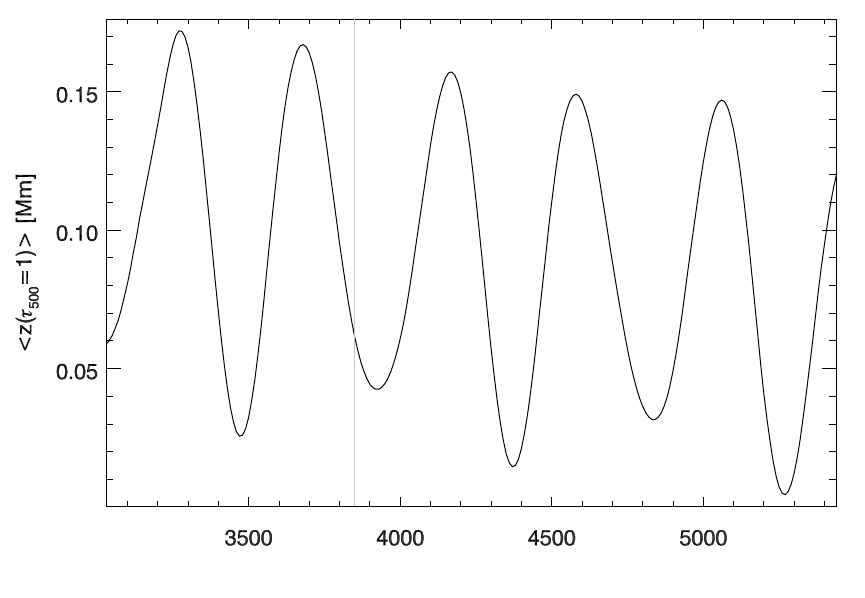}
\caption{Average height of $\tau_{500}=1$ as function of time.
The start of the published sequence of snapshots is indicated at $t=3850$\,s as a grey line.
\label{fig:ztau51}}
\end{figure}

The height scale in the simulation is only approximately normalised to have a zero-point at optical depth unity at 500 nm 
(the usual zero-point of height-scales). Since there are oscillations in the simulation, the average height of $\tau_{500}=1$
varies in time with an amplitude of 60 km and a mean of 89 km, see Figure~\ref{fig:ztau51}.

\subsection{Temperature structure}\label{subsec:temp}

The Joule heating caused by the braiding of the magnetic field from convective motions results in an increased temperature in the
chromosphere and corona in the simulation. Additional heating comes from viscous dissipation. 
In this section we illustrate the temperature distributions found in the simulation but the detailed analysis of the
energy balance is outside the scope of this paper.

\begin{figure}[htbp]
\includegraphics[width=\columnwidth]{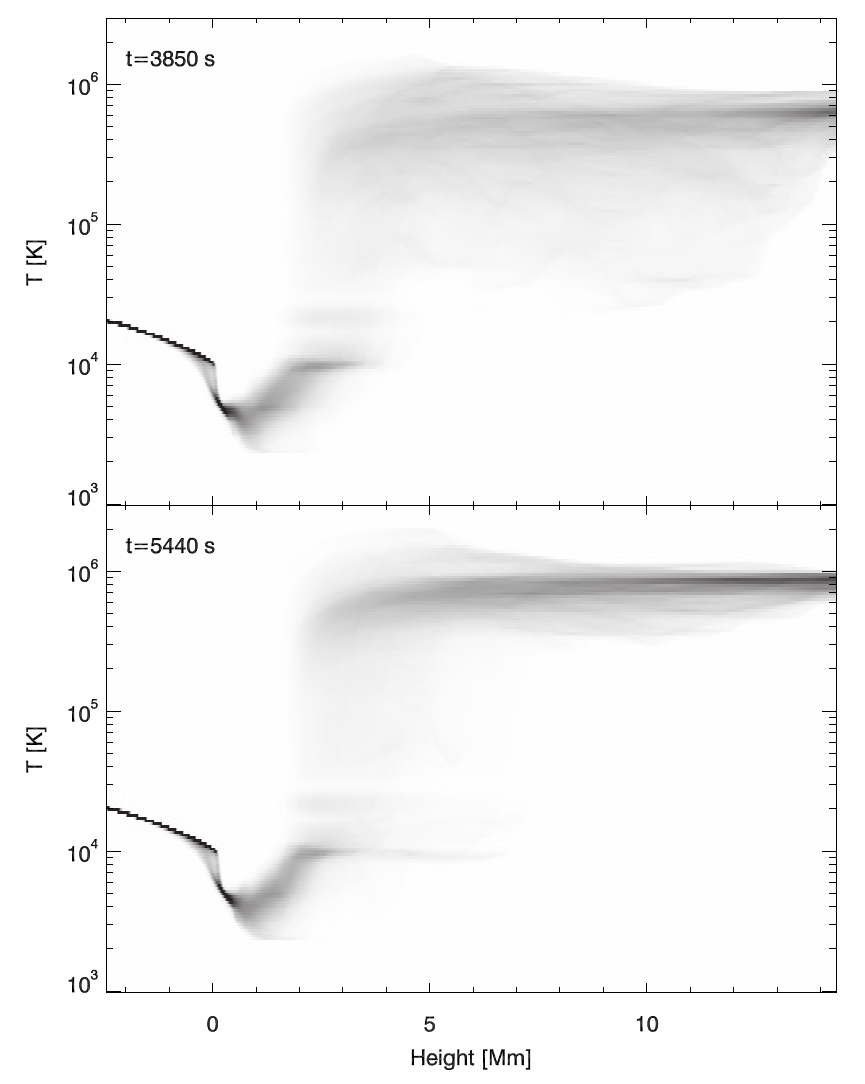}
\caption{Probability density function (PDF) of the temperature as function of height at $t=3850$\,s (upper panel)
and at $t=5440$\,s (lower panel). 
Note the logarithmic temperature scale.
\label{fig:tpdfz}}
\end{figure}

The probability \edt{density} function (PDF) of temperature as function of height at $t=3850$\,s 
is shown in the upper panel of Figure~\ref{fig:tpdfz}.
The spread in temperature at a given height is very small in the deep photosphere and increases in the \edt{subsurface layers}, 
where we have hot granular upflows and cool intergranular downdrafts. There is a pronounced drop in temperature around
$z=0$  and at $z=0.2$\,Mm the temperature is restricted to a range between 4500 and 5500\,K. Further up, there are both
higher and lower temperatures with an average steady increase in the chromosphere up to a height of 2\,Mm. Between
2 and 4\,Mm height we encounter both chromospheric temperatures around $10^4$\,K and transition-region to coronal 
temperatures up to $10^6$\,K. From 4--14\,Mm height we have temperatures up to slightly above $10^6$\,K.
There is a lower limit of 2400\,K set by an artificial heating term that sets in as soon as the temperature drops below
2500\,K. This is necessary in order to prevent the temperature from dropping to very low values in areas of rapid expansion (\eg\
caused by the emergence of magnetic loops), see \citet{2011A&A...530A.124L} for a discussion. There are relatively few points
in the simulation box that are affected by this artificial limit in temperature. The bands of increased probability at temperatures of
\edt{10}\,kK and \edt{20}\,kK are caused by the ionization of helium that is treated in LTE in the current simulation, 
see  \citet{2014ApJ...784...30G} for a discussion of non-equilibrium effects of helium ionization.
At the end of the simulation run, at $t=5440$\,s, the distribution is rather similar to the situation at $t=3850$\,s in the chromosphere, but the corona has been
further heated such that the regions with temperatures below 300\,kK above a height of 5\,Mm are now gone, with the exception of an extended helium ionization region at 10\,kK.

\begin{figure}[t]
\includegraphics[width=\columnwidth]{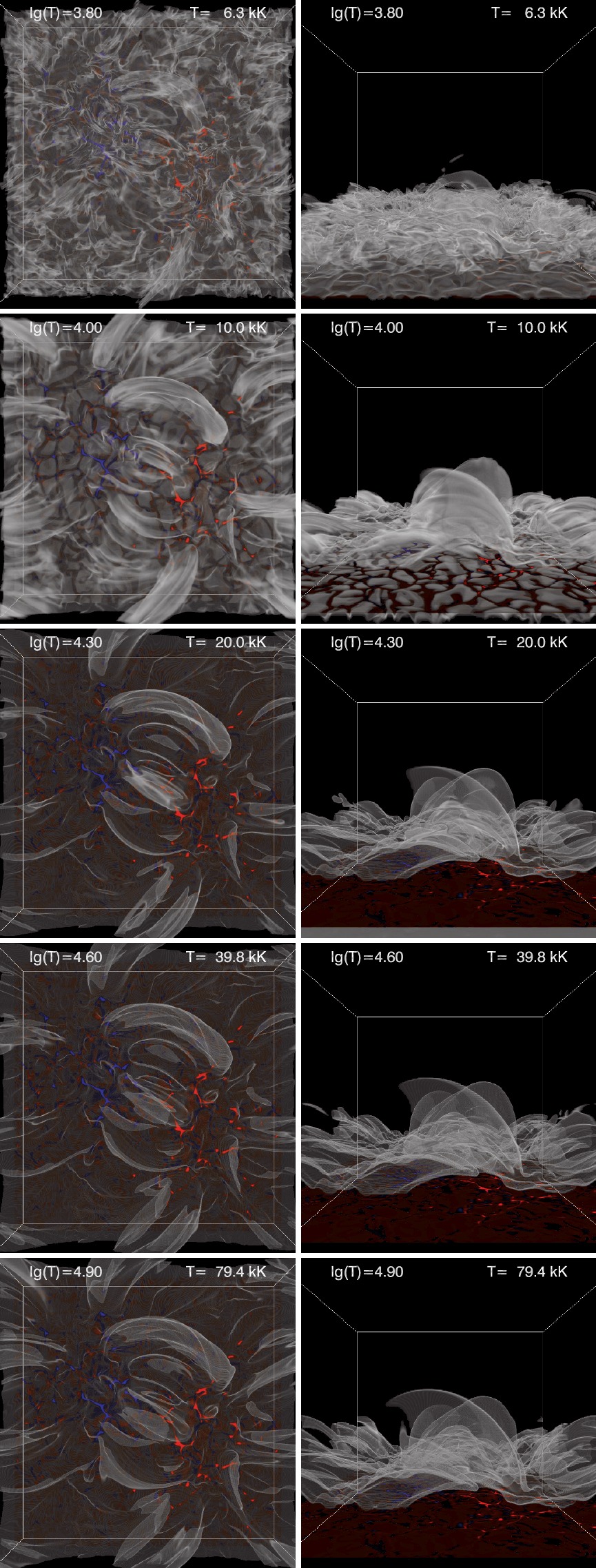}
\caption{Volume rendering of the temperature distribution at $t=5440$\,s viewed from the top (left) and side (right). $B_z$ at $z=0$ with
positive (red) and negative (blue) polarity. The Moir\'e patterns are artefacts of the volume visualisation.
\label{fig:lgta}}
\end{figure}
\begin{figure}[t]
\includegraphics[width=\columnwidth]{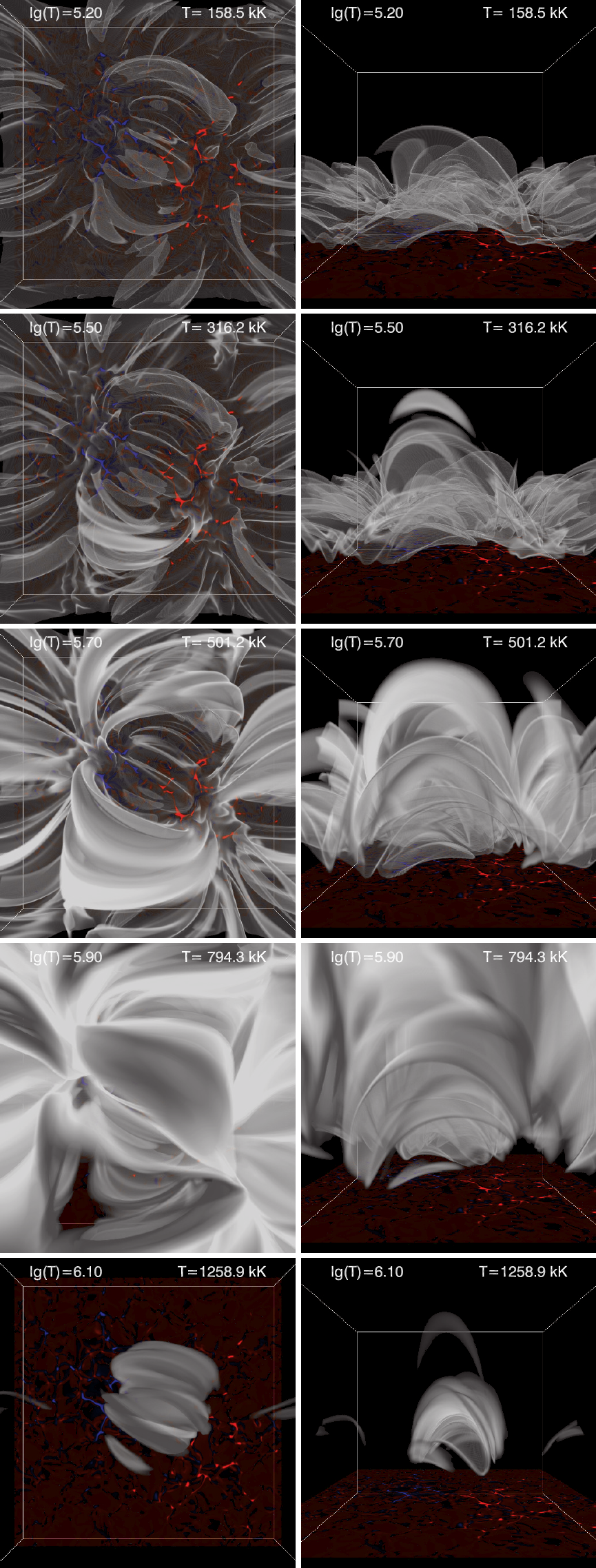}
\caption{Same as Figure~\ref{fig:lgta} for $\log_{10} T=5.2-6.1$
\label{fig:lgtb}}
\end{figure}
As is obvious from Figure~\ref{fig:tpdfz}, the temperature is not a single valued function of height; there is a large spread of temperatures
at most heights. Figures~\ref{fig:lgta}--\ref{fig:lgtb} show the spatial distribution of the plasma at different temperatures. Each panel 
shows the distribution of plasma at a given temperature with a triangular shaped weighting centred on a given logarithmic temperature
with a range of $\pm 0.05$ in the logarithm. 
Note that there is no weighting with density (as would be \edt{appropriate} for an optically thin
spectral line with a given formation temperature).

At a temperature of 6.3\,kK we already see low lying loop structures connecting magnetic field of opposite polarities. There 
is a multitude of
these low lying, short loops but much of the plasma at that temperature is also distributed in structures that are less loop-like.
At 10\,kK most of the lowest lying loops have disappeared and we have fewer, more pronounced loops that reach higher.
At higher temperatures we basically see the same loops, all the way up to 316\,kK ($\log_{10}(T)=5.5$) when a new set of hotter, higher lying loops start to appear. 
These loops dominate up to about 1\,MK. The maximum temperature in this simulation is 2.2\,MK and this hottest plasma 
is located in loops that do not reach as high as the loops with temperatures up to 1\,MK.

The lower lying loops with temperatures below 300\,kK evolve on shorter timescales than the hotter loops and 
give rise to the "Unresolved Fine Structure" (UFS) loops discussed in \citet{2014Sci...346E.315H}.

\subsection{Ionization balance}\label{subsec:ionbal}
\begin{figure}[htbp]
\includegraphics[width=\columnwidth]{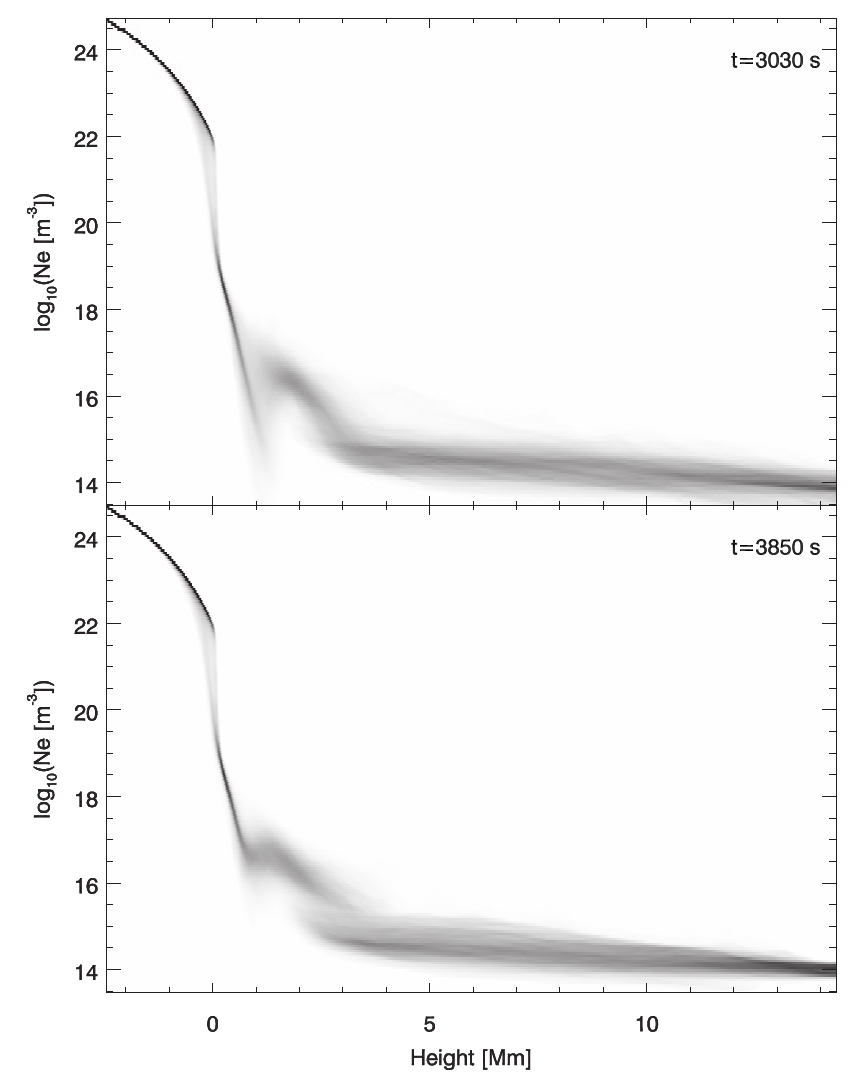}
\caption{Probability density function (PDF) of the electron density as function of height at $t=3030$\,s (upper panel)
and at $t=3850$\,s (lower panel). 
Note the logarithmic electron density scale.
\label{fig:nepdfz}}
\end{figure}

The simulation includes the effects of non-equilibrium ionization of hydrogen, see
\citet{2007A&A...473..625L}. 
Figure~\ref{fig:nepdfz} shows the electron density as a function of height for two times, just 10\,s after the non-equilibrium
ionization of hydrogen was switched on at $t=3030$\,s and for the first published snapshot, at $t=3850$\,s.
The non-equilibrium hydrogen ionization leads to much higher electron density in the cool pockets at 0.5--2\,Mm and
also higher electron densities up to 3.5\,Mm height. 
There is not much change with time of this probability density function after $t=3850$\,s.

\subsection{Velocity field}\label{subsec:vel}
Spectral lines are normally observed to be broader than what the thermal broadening of the opacity profile would give. An extra free parameter, 
called microturbulence, is often introduced in 1D semi-empirical models to account for this broadening. The "micro" in the name comes from 
the fact that this parameter is introduced as an extra broadening of the opacity profile, acting in exactly the same way as thermal broadening. This would
be a physically correct description in the limit of zero length-scale for the velocity field. It is also often necessary to introduce a second free parameter
to account for the observed lineshape. This is called macroturbulence and is equivalent to a Gaussian convolution of the emergent intensity profile
(rather than a convolution of the opacity profile as is the case for microturbulence). 
Realistic 3D radiation hydrodynamic simulations of the solar photosphere give line profiles that
are close to the observed profiles without the addition of extra free parameters --- the non-thermal broadening comes from Doppler shifts arising
from the convective flows and oscillations \citep[\eg][]{2000A&A...359..729A}.

Also spectral lines formed in the outer atmosphere are broader than what thermal broadening alone predicts. 
The nature of this non-thermal broadening in the outer atmosphere 
is still unclear, but the presence of strong shocks 
\citep{1992ApJ...397L..59C,1997ApJ...481..500C,2015ApJ...799L..12D}, 
torsional motions \citep{2014Sci...346D.315D}, 
and Alfv\'en wave turbulence \citep{2011ApJ...736....3V}
are some of the candidates.

\begin{figure}[htbp]
\includegraphics[width=\columnwidth]{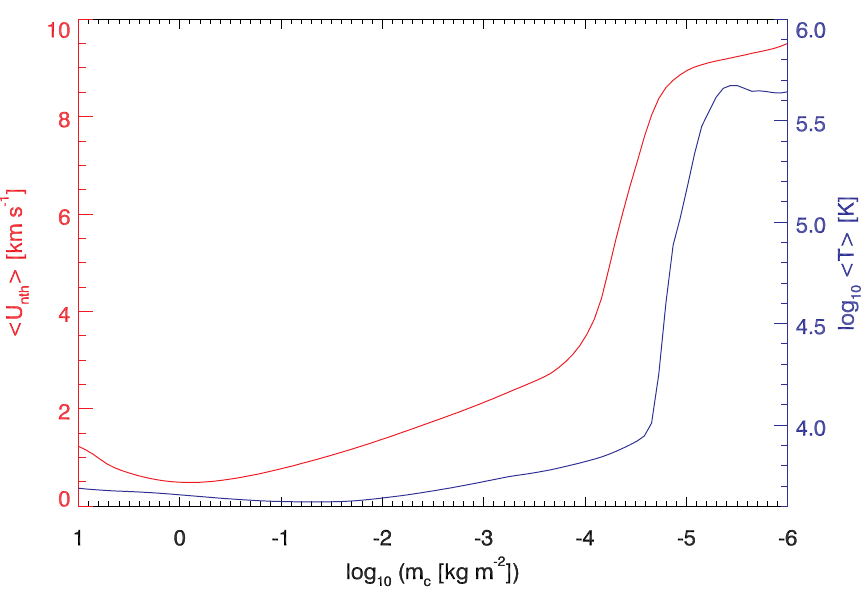}
\caption{Average non-thermal velocity calculated over $\pm 1$ dex in $\log_{10}$(column mass) 
(red, left scale) and average temperature (blue, right scale) as function of logarithmic column mass for
the simulation snapshot at $t=3850$\,s. 
\label{fig:uznthma}}
\end{figure}

There is no simple way to characterize the macroscopic velocities in the simulation that give rise to non-thermal broadening. 
The effect of a given velocity field on the spectral line width depends on whether the spectral line is optically thick or optically thin, where in the
atmosphere the line is formed and the width of the contribution function to intensity. 
One possible way of quantifying the velocity field is the
standard deviation of the vertical velocity over a given height range as function of height and horizontal position.
At each column of the simulation box at $t=3850$\,s we calculate the column mass scale (which is more closely
related to line formation quantities like optical depth than
a geometric height) 
and take the standard deviation of the vertical velocity over 
a range of $\pm 1$ dex in $\log_{10}$(column mass). 
We multiply the standard deviation by the square root of 
two in order to get a quantity that can be directly compared 
with microturbulence and
call this quantity non-thermal 
velocity, $U_{\rm nth}$. 
The average over planes of constant column mass is shown in Figure~\ref{fig:uznthma}. 
The temperature averaged over planes of constant column mass is also shown.
The transition region is situated around $\log_{10} (m_c \mathrm{[kg\,m}^{-2}\mathrm{]})=-5$ (equivalent to a logarithmic value of $-6$ in cgs units).
The average non-thermal velocity rises steadily from 0.5\,\kms\ at zero logarithmic column mass to 3.5\,\kms\ at $\log_{10} (m_c)=-4$. Through
the transition region, the average non-thermal velocity in the simulation rises to 9\,\kms. These values are quite a bit smaller than are needed
to explain the non-thermal broadening of optically thin spectral lines formed in the chromosphere 
\citep[4-8\,\kms;][]{2015ApJ...809L..30C}.
and lower transition region
\citep[$\approx 20$\,\kms;][]{2015ApJ...799L..12D}.

Preliminary results from simulations run at higher spatial resolution (horizontal grid size of 31\,km instead of 48\,km) indicate that part of the 
explanation may be the limited numerical resolution of the current simulation: in the 31\,km simulation the non-thermal velocity at 
$\log_{10} (m_c)=-3$ increased from 2.1\,\kms\ to 3.4\,\kms, the rapid increase in non-thermal velocity happens already at 
$\log_{10} (m_c)=-3.5$ and the value in the transition region increases to 15\,\kms.

\section{Data access}\label{sec:data}
The full simulation cubes with all variables as function of grid
position are available from the Hinode Science Data Centre Europe
(http://www.sdc.uio.no/search/simulations).

Each timestep saved to file is called a snapshot and they are numbered from
$t=0$ with 10\,s of solar time separating each snapshot. The first
published snapshot is snapshot 385 at $t=3850$\,s, which is 830\,s after
the switch on of the non-equilibrium hydrogen ionization when the
initial startup effects have largely disappeared. The last snapshot is
at $t=5440$\,s giving a timespan of 1590\,s for the published
simulation. 

All files are in FITS format with a format similar to IRIS level 2 data: 3D cubes of data ($x$,$y$,$z$) with one variable per file. 
The $x$- and $y$-grids are equidistant and can be generated using the standard FITS keywords while the $z$-grid is 
non-uniform and is therefore given in a FITS extension.

The file names are of the form {\tt BIFROST\_en024048\_hion\_<var>\_<snap>.fits} where the runname {\tt en024048\_hion} comes from
"enhanced network", 24\,Mm horizontal size, 48\,km horizontal grid-spacing and {\tt hion} because the simulation includes
non-equilibrium ionization of hydrogen. {\tt <var>} is the variable name, listed in Table~\ref{tab:variables}, and {\tt <snap>} is the snapshot
number.

\begin{table}[h]
\caption{Available variables}
\begin{tabular}{ll}
\hline\hline
variable & explanation \\
\hline
{\tt lgr} & $\log_{10}$(mass density)\\
{\tt ux} & bulk velocity in $x$\\
{\tt uy} & bulk velocity in $y$\\
{\tt uz} & bulk velocity in $z$\\
{\tt lge} & $\log_{10}$(internal energy)\\
{\tt bx} & magnetic field strength in $x$\\
{\tt by} & magnetic field strength in $y$\\
{\tt bz} & magnetic field strength in $z$\\
{\tt lgne} & $\log_{10}$(electron density)\\
{\tt lgp} & $\log_{10}$(gas pressure)\\
{\tt lgtg} & $\log_{10}$(temperature)\\
\hline
\end{tabular}
\label{tab:variables}
\end{table}

All variables are cell centred on a right-handed system with $z$ increasing upwards.  
Index runs the same way as the axis which means that $z$[1] is at the bottom and $z$[nz] at the top. 
Note that this is different from the original \bifrost\ files.

All units are SI units and given in FITS keywords (Mm, m/s, kg\,m/s, T, W/m$^3$, nm, etc). 
Specifically this means that magnetic field strength is given in Tesla (1 T=$10^4$ G).

Metadata is given in the FITS header. This data release is part of the \iris\ project and an explanation of the FITS keywords is given in
{\it IRIS Technical Note 33} (the {\it IRIS Technical Note} series is available from http://iris.lmsal.com). Software to analyse the simulation data is provided in 
{\it SolarSoft (http://www.lmsal.com/solarsoft)} with descriptions in {\it IRIS Technical Note 34}. Synthetic observables are 
also made publicly available at the Hinode Science Data Centre Europe; so far only the spectrum around the \mgiihk\ lines but
more will follow, see {\it IRIS Technical Note 35}. 

Papers published based on the simulation presented here should cite both the code description paper
\citep{2011A&A...531A.154G} and this paper.

\section{Discussion and Conclusions}\label{sec:discussion}

We have presented the main characteristics of a \bifrost\ simulation aimed at the study of the outer solar atmosphere. The
main free parameter is the initial magnetic field configuration. The field configuration of the current simulation, named {\it en024048\_hion}, is
characterised by two opposite polarity patches separated by some 8\,Mm in a box of horizontal extent 24\,Mm $\times$ 24\,Mm.

It is important to take into account the characteristics when analysing the simulation or synthetic observables derived from it. The
major caveats presented above are:
\begin{itemize}
\item The opacities \edt{ and abundances} are from old tables, basically from \citet{Gustafsson1973}; \edt{\citet{1975A&A....42..407G}}, in order to be compatible with earlier deep convection simulations. 
These opacities \edt{ and abundances} are not ideal for comparison of synthetic observables with detailed photospheric intensities.
\item The effective temperature is not specified in the simulation and is only set by specifying the entropy of the incoming fluid at the bottom boundary. 
The relaxation to a given effective temperature is a very slow process and in the {\it en024048\_hion} simulation the effective temperature is typically lower than that of the Sun, see Section~\ref{subsec:photosphere}.
\item There are major oscillations in the simulations, see Section~\ref{subsec:oscillations}.
\item The height scale is only approximately normalised to have a zero-point close to optical depth unity at 500 nm (the usual zero-point of height-scales). 
Since there are oscillations in the simulation, the average height of $\tau_{500}=1$ varies in time, see Section~\ref{subsec:oscillations}
\item The published data have all variables specified at the same location (cell-centres) instead of being on a staggered grid as in the original simulation. 
This means that the variables that originally are not given at cell-centres (velocities and magnetic field strength) 
have been interpolated to cell-centres with the same high-order interpolation scheme as used in \bifrost. 
This introduces interpolation noise, in particular the divergence of B is no longer zero to the machine accuracy as is the case for the original data.
\end{itemize}

The paper series on "The formation of \iris\ Diagnostics" (see Section~\ref{sec:introduction}) contains several comparisons of synthetic observables from this simulation with observations.
It is clear from these comparisons that the simulation lacks 
important physics,
even for the quiet sun. In particular chromospheric spectral lines synthesised from the simulation tend to be too weak and too narrow. The comparisons
indicate that the simulation has too small amplitude mass motions at small spatial scales (the "non-thermal broadening" of spectral lines is too small) and
too little plasma at chromospheric temperatures. However, the parameter space exhibited by the simulation seems to cover typical chromospheric conditions (albeit
not in the right proportions) and we hope the simulation sequence published here can serve as a useful laboratory to further our understanding of the
outer solar atmosphere.

\begin{acknowledgements}
The research leading to these results has received funding from the European Research
Council under the European Union's Seventh Framework Programme (FP7/2007-2013) /
ERC Grant agreement n$^o$ 291058.
This research was supported by the Research Council of Norway through
the grant ``Solar Atmospheric Modelling'' and 
through grants of computing time from the Programme for Supercomputing
and through computing project s1061 from the High End Computing (HEC) division of NASA.
We acknowledge PRACE for awarding us access to resource HERMIT based in Germany at GCS
in HLRS. B.D.P. was supported by NASA contract NNG09FA40C (IRIS).
IRIS is a NASA small explorer developed and operated by LMSAL with mission operations executed at NASA Ames and major contributions to downlink communications funded by ESA and the Norwegian Space Centre.
CHIANTI is a collaborative project involving George Mason University, the University of Michigan (USA) and the University of Cambridge (UK).
We have used VAPOR \citep{vapor1,vapor2} extensively for visualisations.
\end{acknowledgements}

\bibliographystyle{aa}

\end{document}